# Performance Evaluation of Video Communications over 4G Network

Gaurav Pande
Department of MCA,
RK Goel Institute of Technology
Ghaziabad, India
`reachgauravpande@gmail.com`

**Abstract.** With exponential increase in the volumes of video traffic in cellular networks, there is an increasing need for optimizing the quality of video delivery. 4G networks (Long Term Evolution – Advanced or LTE-A) are being introduced in many countries worldwide, which allow a downlink speed of upto 1 Gbps and uplink of 100 Mbps over a single base station. In this paper, we characterize the performance of LTE-A physical layer in terms of transmitted video quality when the channel conditions and LTE settings are varied. We test the performance achieved as the channel quality is changed and HARQ features are enabled in physical layer. Blocking and blurring metrics were used to model image quality.

## 1    Introduction

Mobile video is quickly becoming a mass consumer phenomenon, much as digital photos were earlier in the smartphone adoption cycle. There is a tectonic shift to on-demand and on-the-go video services. Video streaming services such as Netflix and Youtube are generating large volumes of traffic and revenue (3 billion $ for Netflix in 2011).

Moreover, the cellular networks are migrating towards a new technology 4G LTE which is 10x faster than 3G. Boosted bandwidth allows for quicker uploads and better streaming quality. LTE-A targets to achieve peak uplink and downlink data rates of 500 Mbps and 1 Gbps, respectively, for low-speed UEs and around 100 Mbps for those with higher mobilities. It accommodates the next generation of telecommunication services such as realtime high-definition video streaming, mobile HDTV and high quality video conferencing. In LTE-A systems, the bandwidths in both uplink and downlink can go upto 100 MHz, which is achieved by Carrier Aggregation (CA) or aggregation of individual Component Carriers

There have been volumes of research in different aspects of multimedia systems, to enhance the performance of mobile video. Hardware architectures have been proposed for mobile video compression and encryption [9-11]. Efficient video transmission in cellular low bandwidth scenarios has also been discussed [12-16]. With 4G LTE becoming popular, we attempt to understand the features of LTE which help in high quality video transmission. The major features that distinguish LTE from 3G technologies at the air-interface are Orthogonal Frequency Multiple Access

(OFDMA), advanced MIMO technology, and Hybrid Automatic Repeat Request (HARQ). In addition LTE uses flat-IP architecture for the core network. LTE uses OFDMA in the downlink (DL) for efficient multiple-access and for countering multi-path frequency selective fading. OFDMA divides the available channel into number of sub-carriers and is naturally suitable for scalable bandwidth allocation by varying the FFT (Fast Fourier Transform) size. LTE uses SC-FDMA (Single Carrier FDMA) [6] in the uplink (UL) to obtain a low peak-to-average-power ratio (PAPR). In this paper, we concentrate only on the DL. LTE supports a full range of multiple antenna transmission techniques including transmit diversity (TD) [2], spatial multiplexing (SM) [3], and closed-loop eigen-beamforming [4] that are suited for different objectives. Transmit diversity is used for obtaining reliable transmissions and is achieved by using Space Frequency Block Codes [5] in LTE. SM is used for obtaining enhanced throughput and is achieved by using layered space time codes [3]. Eigen-beamforming also is used to improve reliability of transmissions when accurate channel state information is available. Traditional link adaptation techniques used channel quality information only to adapt the MCS level used for transmissions.

In this paper, we use the LTE Physical layer simulator presented earlier in [7,8]. The features of this simulator correspond to a practical LTE station. It is developed in Matlab. There has been impending interest in studying the quality of video transmitted over mobile networks. Many researchers have discussed this problem in great length [16-20]. The main motivation in studying the quality of video delivered over mobile devices is as follows:

1. By understanding the video quality delivered to end user, we can effectively quantize the performance of video application. This is not possible by mere measurements of network QoS parameters. Studying video quality helps us to closely understand how the end users will perceive the transmitted video.

2. With exponential increase in mobile traffic, the capacity of cellular networks will be saturated very quickly. According to Peter Rysavy, a wireless analyst in USA, mobile broadband will surpass the spectrum available in mid-2013 [21]. If we can measure the perceived video quality by end users, we can optimize the quality so as to provision more number of users in same bandwidth.

3. Video quality is subjective and some small distortions may not be visible. This opens a new area of perceptual optimization where network resource allocation protocols are designed to provide video experience measurably improved under perceptual models of human eye [16].

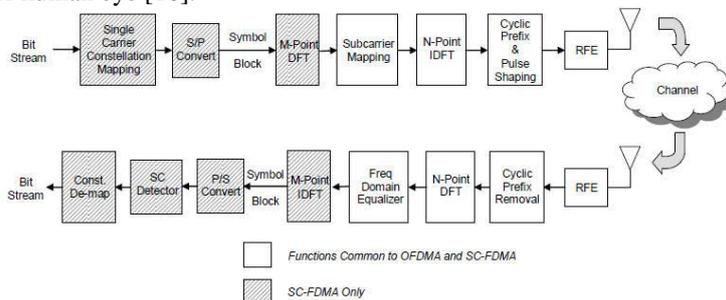

Fig. 1. **Overall block diagram of physical layer in LTE-A**

To this end, in this paper, we present an evaluation of video quality in video streaming services in LTE-A.

## 2   LTE PHY details

In this section, we detail about the various parameters in LTE-A, their meaning and significance and also explain the choices we used for implementation for uplink transmission. OFDM is a well-known modulation technique, but is rather novel in cellular applications. LTE uses a rate-1/3 Parallel Concatenated Convolutional Code (PCCC) consisting of two identical 8-state rate-½ convolutional encoders connected parallel using an internal interleaver.  Viterbi decoding of turbo codes is complex due to the large number of states involved in a concatenated trellis. So an iterative MAP (Maximum A Posteriori) detector based algorithm [6] is used as a practical alternative decoding scheme. In the DL, LTE uses asynchronous and adaptive HARQ mechanism. The schedule of the HARQ transmissions is not pre-declared to the UE. This gives the eNodeB flexibility in scheduling according to priorities and available resource. LTE uses Incremental Redundancy (IR) HARQ as opposed to chase combining. LTE supports up to four redundancy versions for IR HARQ (re)transmissions denoted by rvidx = 0,1,2 and 3. In each version, a part of rate-1/3 turbo-encoded data is transmitted dependent on rvidx value.

The rate matching converts the rate-1/3 output from the turbo encoder into the target coding rate. This is done by a block consisting of a three sub-block interleavers, a circular buffer, and a bit-selection block [21]. The number of bits selected depends on the target coding rate. The start point (or offset) of the selected bits is determined by the HARQ redundancy version rvidx. OFDMA LTE uses OFDMA for DL access. The available frequency is divided into sub-carriers of 15 kHz bandwidth. LTE specific OFDMA parameters are listed are listed in Table 1 [22].

A tapped delay line model is used to model multipath frequency selective channel between transmit antenna and receive antenna [7]. The metrics used for this work are: PSNR or Peak Signal to Noise Ratio, SSIM or Structural Similarity Metric [23-24], blocking [25-28]and blurring[29].

## 3   EVALUATION

To test the comparative performance of LTE Physical layer in our simulations, we bypasss the higher layers such as transport protocol, IP layer, and Network layer. Rather, we directly connect application layer data to the physical layer. Since we are interested in evaluating the comparative performance of PHY level parameters, this change in configuration doesn't matter. The Simulink model proposed earlier by earlier researchers [7,8] was used for the simulations. The quality of video was evaluated in terms of blocking metrics by implementing the reference implementation of most cited work[28].

We choose three video samples – Akiyo video, Rhino video and harbor video to maintain diversity in the sample set.

### 3.1 Experiments

To perform this set of experiments, we took a constant value of other parameters such as modulation (kept at 32 QAM) and code-rate (2/3), HARQ (kept at 1) and MIMO (disabled). We then changed the channel SNR (EbNo) value from 0 to 20 db. We observe a very poor performance for low SNR (high noise) where practically the entire image has severe blocking and unacceptable quality. As the SNR goes above 18, we observe good quality video streaming.

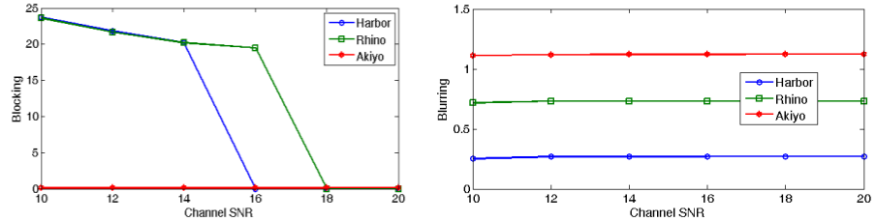

**Fig. 2.** Blocking and blurring measurements for Akiyo video sequence as a function of channel variations. The modulation and code rate and HARQ/MIMO (off) parameters are kept constant to take these observations.

Figure 3 shows the blocking and blurring values measured for the three test video sequences – Harbor, Rhino and Akiyo respectively. It can be clearly seen how blurring metric is agnostic to network channel conditions and the value remains practically the same throughout. The Blocking metric performance was measured in logarithm scale as the numerical values are too large. It can be seen that different videos require different bit-rates and hence different channel conditions for satisfactory transmission. In our case, Akiyo video can be transmitted fairly properly even at 10 db while other videos require 16 or 18 db SNR. We hope that this performance can be improved by using HARQ technique for Harbor and Rhino videos. Next, we change the HARQ retransmission rates while keeping the other parameters constant. We do this for Harbor video sequence and report the observations in Figure 4. This can be confirmed here, that blurring metrics don't provide any significant information about network degradations. There is no consistency in the results for varying values of HARQ retransmissions. The blocking metrics give consistent performance, we can see that increasing the number of retransmissions reduces the blocking and increases video quality.

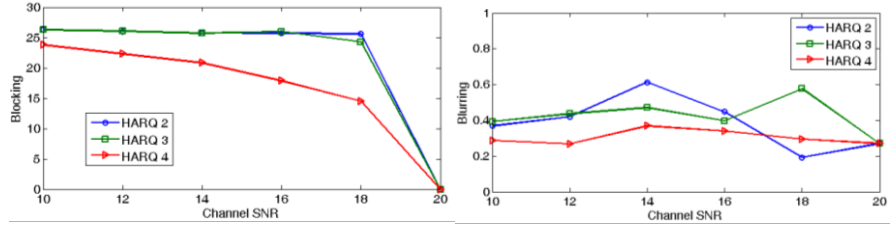

**Fig. 3.** Variation in blocking and blurring with changes in HARQ retransmission options for Harbor video sequence

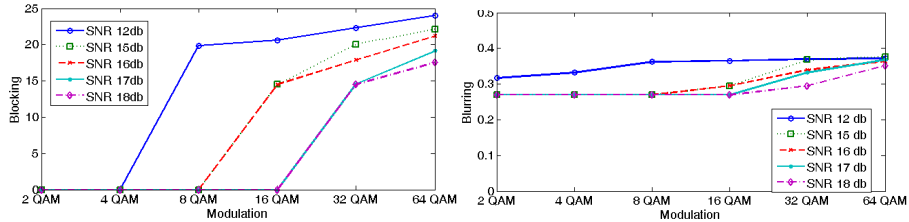

**Fig. 4.** Variation in blocking and blurring with changes in modulation scheme used in LTE PHY for different channel conditions

We next observe the change in video quality with changes in channel conditions. In the LTE Physical layer model, we input the channel SNR (EbNo) value to indicate the channel conditions. Depending on this and the modulation/ code rate chosen, we obtain the exact error rate for the system. Therefore, we plot this time as a function of modulation rates, keeping the code rate fixed to 2/3 and HARQ set to 4. As the SNR increases, the channel improves and it can be seen that higher modulation rates can be used while maintaining lower blocking values. In this case, blurring measurements are consistent, however, the difference between them is small.

## 4      Discussions & Conclusions

In this paper, we examined the performance of LTE physical layer with respect to video delivery. We used two popular image distortion measurement metrics, namely blocking and blurring. We find that blurring artifacts are not very correlated with the image quality, particularly when the distortions are large. Blocking measure gives results consistent with our theoretical understandings.

## 5      References

[1] R Sivaraj, A Pande, K Zeng, K Govindan, P Mohapatra, "Edge-prioritized channel-and traffic-aware uplink Carrier Aggregation in LTE-advanced systems" WOWMOM 2012
[2] S. M. Alamouti, "A simple transmit diversity technique for wireless communications," IEEE Journal on Selected Areas in Communications, vol. 16, pp. 1451-1458, 1998.
[3] G. J. Foschini, "Layered space-time architecture for wireless communication in a fading environment when using multi-element antennas," Bell Labs Technical Journal, vol. 1, pp. 41-59, 1996.
[4] E. Sengul, E. Akay, and E. Ayanoglu, "Diversity analysis of single and multiple beamforming," IEEE Transactions on Communications, vol. 54, pp. 990-993, 2006.
[5] H. Bolcskei and A. J. Paulraj, "Space-frequency coded broadband OFDM systems," in Proc. IEEE WCNC '00, vol. 1, pp. 1-6 vol.1, 2000.
[6] H. G. Myung, J. Lim, and D. J. Goodman, "Single carrier FDMA for uplink wireless transmission," IEEE Vehicular Technology Magazine,vol. 1, pp. 30-38, 2006.
[7] V. Ramamurthi, and C. W. Peng, "Mobility based MIMO Link adaptation in LTE-Advanced Cellular networks." Broadband, Wireless Computing, Communication and Applications (BWCCA), 2010 International Conference on. IEEE, 2010.
[8] A. Pande, V. Ramamurthi, and P. Mohapatra. "Quality-oriented Video delivery over LTE using Adaptive Modulation and Coding." Global Telecommunications Conference (GLOBECOM 2011), 2011 IEEE. IEEE, 2011.

[9] A. Pande, and J. Zambreno, "An Efficient Hardware Architecture for Multimedia Encryption and Authentication Using the Discrete Wavelet Transform." VLSI, 2009. ISVLSI'09. IEEE Computer Society Annual Symposium on. IEEE, 2009.

[10] A. Pande, and J. Zambreno, "Poly-DWT: Polymorphic wavelet hardware support for dynamic image compression." ACM Transactions on Embedded Computing Systems (TECS) 11.1 (2012): 6.

[11] A. Pande, and J. Zambreno,"A reconfigurable architecture for secure multimedia delivery." VLSI Design, 2010. VLSID'10. 23rd International Conference on. IEEE, 2010.

[12] A. Pande, J. Zambreno and P. Mohapatra. "Hardware architecture for simultaneous arithmetic coding and encryption." International Conference on Engineering of Reconfigurable Systems and Algorithms. 2011.

[13] A. Sood, D. Sarthi, A. Pande, and A. Mittal, "A novel rate-scalable multimedia service for E-learning videos using content based wavelet compression". In IEEE India Conference (INDICON), 2006 (pp. 1-6).

[14] A. Pande, A. Mittal, A. Verma and P. Kumar, " Meeting real-time requirements for a low bitrate multimedia encoding framework". In Electro/Information Technology, 2008. EIT 2008. IEEE International Conference on (pp. 258-262). IEEE.

[15] E. Baik, A. Pande, and P. Mohapatra. "Cross-layer Coordination for Efficient Contents Delivery in LTE eMBMS Traffic." IEEE MASS 2012

[16] P. McDonagh, C. Vallati, A. Pande, P. Mohapatra, P. Perry, and E. Mingozzi, "Quality-Oriented Scalable Video Delivery using H. 264 SVC on an LTE Network", International Conference on Wireless Personal Multimedia Communications(WPMC) , 2011.

[17] A. Moorthy, L. Choi, A. Bovik, G. D. Veciana, "Video quality assessment on mobile devices: Subjective, behavioral and objective studies" 2012.

[18] A. Mittal, R. Soundararajan, and A. Bovik. "Making a 'Completely Blind'Image Quality Analyzer." (2012): 1-1. IEEE Signal Processing Letters

[19] A. Mittal, A. Moorthy, and A. Bovik. "No-Reference Image Quality Assessment in the Spatial Domain." IEEE Transactions on Image Processing (2012): 1-1.

[20] A. J. Chan, A. Pande, E. Baik, and P. Mohapatra, "Temporal quality assessment for mobile videos". In Proceedings ACM Mobicom 2012 (pp. 221-232).

[21] 3GPP TS 36.212: "Evolved Universal Terrestrial Radio Access (EULTRA); Multiplexing and Channel Coding," 2008.

[22] 3GPP TS 36.211: "Evolved Universal Terrestrial Radio Access (EULTRA); Physical channels and modulation," 2008.

[23] Z. Wang, A. Bovik, H R Sheikh, and E. P. Simoncelli. "Image quality assessment: From error visibility to structural similarity." Image Processing, IEEE Transactions on 13, no. 4 (2004): 600-612.

[24] Z. Wang, L. Lu, and A. Bovik. "Video quality assessment based on structural distortion measurement." Signal processing: Image communication 19, no. 2 (2004): 121-132.

[25] H. R. Wu, and M. Yuen. "A generalized block-edge impairment metric for video coding." Signal Processing Letters, IEEE 4, no. 11 (1997): 317-320.

[26] D. Wei and A. C. Bovik, "A new metric for blocking artifacts in block transform-coded images and video,"
Technical Report to Southwestern Bell Technology Resources, Inc., Oct. 1998.

[27] Z. Wang, H. R. Sheikh, and A. Bovik. "No-reference perceptual quality assessment of JPEG compressed images." Image Processing. 2002. Proceedings. 2002 International Conference on. Vol. 1. IEEE, 2002.

[28] Z. Wang, A. Bovik, and B. L. Evan. "Blind measurement of blocking artifacts in images." Image Processing, 2000. Proceedings. 2000 International Conference on. Vol. 3. IEEE, 2000.

[29] X. Marichal, M. Wei, and H. Zhang. "Blur determination in the compressed domain using DCT information." Image Processing, 1999. ICIP 99. Proceedings. 1999 International Conference on. Vol. 2. IEEE, 1999.